\documentclass[
 superscriptaddress,
 twocolumn,
 amsmath,amssymb,
 prl,
]{revtex4-2}

\usepackage{graphicx}
\usepackage{dcolumn}
\usepackage{bm}
\usepackage{enumerate}
\usepackage{threeparttable}
\usepackage{multirow}

\begin{document}


\title{Multicolor groups for molecules and solids}

\author{Hai-Yang Ma}
\email{mahaiyang@quantumsc.cn}
\affiliation{Quantum Science Center of Guangdong-Hong Kong-Macao Greater Bay Area, Shenzhen 518045, China}

\author{Shihao Zhang}
\affiliation{School of Physics and Electronics, Hunan University, Changsha 410082, China}

\author{Hu Xu}
\email{xuh@sustech.edu.cn}
\affiliation{Department of Physics, Southern University of Science and Technology, Shenzhen 518055, China}

\author{Shengbai Zhang}
\affiliation{Department of Physics, Applied Physics and Astronomy, Rensselaer Polytechnic Institute, Troy, New York 12180, USA}

\author{Jin-Feng Jia}
\email{jfjia@sjtu.edu.cn}
\affiliation{Quantum Science Center of Guangdong-Hong Kong-Macao Greater Bay Area, Shenzhen 518045, China}
\affiliation{Department of Physics, Southern University of Science and Technology, Shenzhen 518055, China}
\affiliation{Key Laboratory of Artificial Structures and Quantum Control (Ministry of Education), TD Lee institute, School of Physics and Astronomy, Shanghai Jiao Tong University, 800 Dongchuan Road, Shanghai 200240, China}
\affiliation{Hefei National Laboratory, Hefei 230088, China}

\date{\today}

\begin{abstract}
The local magnetic moments of atoms in a molecule or solid can be designated by different colors. Magnetic groups, or 2-color groups, or black-and-white groups 
have been applied in crystallography to classify different magnets. Despite its successes in the past decades, the recent advents of altermagnets and $p$-wave magnets raise new challenges to this long-standing framework, which urges for a new and unified one.
Here we develop a multicolor group classification framework to classify all kinds of molecules and solids, including nonmagnetic materials and magnets with collinear or non-collinear magnetism, and with or without spin-orbit couplings (SOC). This new scheme can unify the classifications of matters into a single framework, including the recently identified altermagnets and $p$-wave magnets. Especially, altermagnetic topological matters and $p$-wave magnets \textbf{with} SOC, can also be diagnosed with multicolor groups, a task which can not be accomplished by magnetic space groups and spin space groups. 
Moreover, insufficiencies and misconceptions of conventional magnetic group classification can be supplemented through this new framework. Multicolor group will serve as a new stage in the symmetry classification of matters.

\end{abstract}

\maketitle


\section[\label{sec.1}]{Introduction}
The spatial symmetry of molecules and solids, including translation, rotation, inversion, and their combinations, is described using point groups and space groups. Simple extensions of these symmetry groups\textemdash the magnetic point group and magnetic space group, which incorporate time-reversal symmetry $\Theta$ into the symmetry operations, can be used to describe and classify different magnetic materials. There are 1651 magnetic space groups, which are partitioned into three (or four) types depending on how time-reversal symmetry is applied to the original space group. A magnetic group, also known as a Shubnikov group, Heesch group, or if no confusion was brought, a dichromatic, 2-color, or simply black-and-white group \cite{shubnikov1964colored,bradley1968magnetic,hahn1983international,lifshitz2004magnetic,litvin2013magnetic,tasci2025computational}. The full enumeration of all magnetic groups has advanced modern condensed matter physics, particularly in the study of altermagnets \cite{vsmejkal2022emerging,PhysRevX.12.031042,mazin2022altermagnetism,ma2021multifunctional,krempasky2024altermagnetic,zhu2024observation,amin2024nanoscale,ma2024altermagnetic,ma2024altermagnetic2,ding2024large,jiang2025metallic,zhang2025crystal}, topological phases \cite{slager2013space,kruthoff2017topological,watanabe2018structure,elcoro2021magnetic}, and symmetry-protected phenomena \cite{gu2009tensor,wen2014symmetry}. However, there are still some insufficiences and misconceptions for the magnetic group. First, magnetic groups are built up without including the spin-orbit-coupling (SOC), this point in general was not clearly indicated. For paramagenets, we can use double groups to account for SOC \cite{tasci2025computational}. While for magnets, generalization from corepresentation theory to magnetic double groups is not so straight forward \cite{bradley1968magnetic}. Second, while the mathematical definitions of the different types of magnetic group are elegant, the physical partitions of different magnets are confused. For example, the type 2 magnetic groups, of which $\Theta$ is not a symmetry operation, usually can describe ferromagnets, while confusingly, it can also describe antiferromagnets, and even for some non-collinear magnets. Third, non-collinear magnets have not been considered in magnetic groups, though we generally also use magnetic groups to designate different non-collinear magnets.

\begin{figure}[tb]
\includegraphics[width=0.48\textwidth]{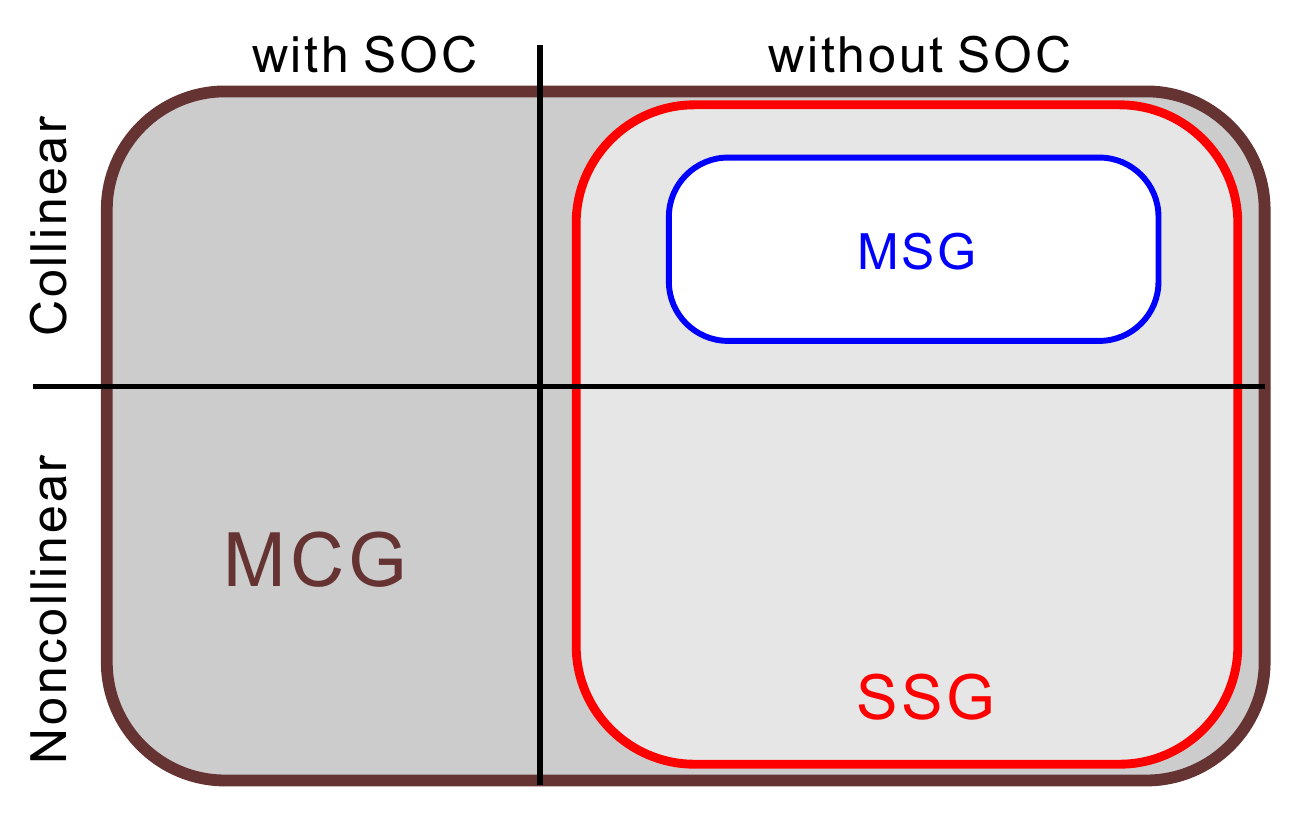}
\caption{\label{fig1} Different partitions of multicolor groups (MCG). The magnetic space groups (MSG) are defined for collinear magnets without including of SOC, which belong to a subpatition of the spin space groups (SSG). Multicolor color groups are applicable for collinear or noncollinear magnets with or without SOC.}
\end{figure}

Recently, the third insufficiency has been partially supplemented by spin space group theory \cite{PhysRevX.12.031042,liu2022spin,jiang2024enumeration,xiao2024spin,chen2024enumeration}. A spin group symmetry operation can be written as $\{R_i||R_{\beta}\}$, where $R_i$ is a symmetry operation in spatial space and $R_{\beta}$ is a symmetry operation in spin space with $R_i\neq R_{\beta}$. Similar to magnetic groups, spin groups are defined in the limit of vanishing SOC, but unlike magnetic groups, their application is restricted to the cases without SOC. This is indeed a theoretical advance that partially removes the confusion. More interestingly, altermagnetism, which has been identified as the third type of fundamental collinear magnetism, was introduced within the workflow of spin space group theory. Nevertheless, the first two issues of the magnetic group remain unscratched. Moreover, altermagnets are generally considered to live in the limit of vanishing SOC \cite{vsmejkal2022emerging,PhysRevX.12.031042,mazin2022altermagnetism}, which can be described by the spin space groups\cite{PhysRevX.12.031042,liu2022spin,jiang2024enumeration,xiao2024spin,chen2024enumeration}. With multicolor groups, we find that \textit{altermagnetism can coexist with SOC}, which means that altermagnets can be intrinsically topological, leading to interesting physics like altermagnetic topological insulator, topological superconductivity, etc \cite{ma2024altermagnetic,ma2021multifunctional,gonzalez2025spin,antonenko2025mirror,liu2025light,wan2025interplay,gonzalez2025model,zhu2025altermagnetic}. In addition, $p$-wave magnets whose colors are related by a half-translation followed by $\Theta$, had recently been predicted and experimentally observed \cite{hellenes2023p,brekke2024minimal,sukhachov2024impurity,ezawa2024topological,maeda2025classification,chakraborty2025highly,soori2025crossed,nagae2025flat,sukhachov2025coexistence,yamada2025gapping,song2025electrical,zhou2025anisotropic}. However, these magnets had not yet been clearly classified. In this work, we perform the classification and we show that more candidate $p$-wave magnets, even with SOC, can be predicted within the multicolor-group framework.

\begin{table*}
\centering
\caption{Multicolor group for molecules and solids. The following abbreviations are used: PM = paramagnets, FM = ferro(ferri)magnets, AM = antiferromagnets, and N = non-collinear magnets. $G_M$ refers to the symmetry group of the magnetic lattice, $G$ refers to the space (sub)group corresponded to the non-magnetic lattice, $H$ refers to a subgroup of $G$, $g'$ refers to an antiunitary symmetry operation, and $g''$ refers to an antiunitary, or a unitary color exchange operation. $\mathcal{P}$ is the color exchange operation of a multicolor group, which is usually a rotation $C_m$, and possibly compound with spatial inversion i, fractional translation $\bm{\tau}_m$ is with respect to the non-magnetic unit cell, and $\sigma$ is a mirror reflection (A collection of such symmetry operations has a group structure isomorphic to $\mathbb{Z}_m$), and $\mathcal{A}$ is a symmetry operation, $\sigma$, $\sigma\bm{\tau}, $i$\Theta$, i$\Theta\bm{\tau}$, or their combinations, which has a group structure isomorphic to $\mathbb{Z}_2$. "/" indicates that this case has not been taken into consideration, since it has been addressed by spin space group \cite{PhysRevX.12.031042,liu2022spin,jiang2024enumeration,xiao2024spin,chen2024enumeration}. }
\label{table3}
\begin{threeparttable}
    \begin{tabular}{c|c|c||c|c}
    \hline
       -color  &  \multicolumn{2}{c||}{without SOC}  & \multicolumn{2}{c}{with SOC} \\
         &     Def. & App. & Def. & App. \\
        \hline
       0 &  $G_M=G\times\mathbb{Z}_2(\Theta)$   & PM & $G_M=G\times\mathbb{Z}_2(\Theta)$ &PM \\
       1  &  $G_M=G$ &  FM & $G_M=H+g''(G-H)$ &FM \\
       2  & $G_M=H+g'(G-H)$ &   AM & $G_M=H+g''(G-H)$ & AM, N \\
       \hline
       3  &    \multicolumn{2}{c||}{/} &  $G_M=H\times\mathbb{Z}_m(\mathcal{P})$ &N\\
       4  &    \multicolumn{2}{c||}{/} &  $G_M=H\times\mathbb{Z}_m(\mathcal{P})$ & N \\
       6  & \multicolumn{2}{c||}{/} & $G_M=H\times\mathbb{Z}_m(\mathcal{P})$ & N \\
       2$\times m\ (m=2, 3, 4, {\rm or}\ 6$) & \multicolumn{2}{c||}{/} & $G_M=H\times\mathbb{Z}_2(\mathcal{A})\times\mathbb{Z}_m(\mathcal{P})$ & N \\
    \hline
       $m$\ ($m>6$ or $m=5$) & $G_M=G\times\mathbb{Z}_m(\bm{\tau}_mC_m)$ & N, crystals & $G_M=H\times\mathbb{Z}_m(\mathcal{P})$ & N, molecules\\
       $2\times m$\ ($m>6$ or $m=5$) & $G_M=G\times\mathbb{Z}_m(\bm{\tau}_mC_m)\times\mathbb{Z}_2(\mathcal{A})$ & N, crystals & $G_M=H\times\mathbb{Z}_2(\mathcal{A})\times\mathbb{Z}_m(\mathcal{P})$ & N, molecules \\
    \hline
    \end{tabular}
    \end{threeparttable}
\end{table*}

Here, we develop a multicolor group \cite{group1999,radaelli2025color} classification framework to address all three insufficiencies of conventional magnetic group classification. Our first step is to clearly separate the cases with and without SOC, collinear and non-collinear magnetism into four partitions. Without SOC, the spatial and spin spaces are independent of each other, which allows us to treat lattice and spin space symmetries separately. This is the starting point of conventional magnetic space groups and spin space groups. Although all real materials have finite SOC, it is still worthwhile to study the cases without SOC since on one hand, they are more diverse and tractable. On the other hand, SOC effects can often be treated as symmetry reductions from the cases without SOC. Despite this, we emphasize that a clear separation, as demonstrated here in Fig.~\ref{fig1}, is needed to avoid confusion. Our next step is to establish the multicolor group theory for each of these cases, and the necessity of separating collinear and non-collinear magnets will manifest during our gradual discussions. The advantage of this classification framework is that it unifies the description of all cases into a single framework through multicolor group theory. We summarize the main results into Table~\ref{table3}. Specifically, we can see from the first 3 rows of Table~\ref{table3} that non-magnetic compounds fall into 0-color groups, ferro/ferrimagnets fall into 1-color groups, antiferromagnets, as usual, fall into 2-color groups. The next 4 rows are for simple $m$-color groups and complex $2\times m$-color groups with $m=3, 4$ or 6, which are applicable to non-collinear magnets. For crystals with translation symmetry, the allowed $m$-color groups generally cannot have $m>6$. The special case are that when SOC is not included, we can have spin-wave constructions, where $m$ can take any integral. For molecules and quasi-crystals, translation is not a symmetry operation, as such the $m$-color point groups are applicable with any integer $m$. These two cases constitute the last 2 rows of Table~\ref{table3}.

\section[\label{sec.2}]{Review of magnetic space groups}
To begin with, we briefly review the definition of the conventional magnetic group. Here, the nonmagnetic point and space group are denoted as $G$, and the magnetic point and space group are denoted as $G_M$. If the antiunitary time-reversal operation, which flips the spin, is denoted as $\Theta$, the three (or four) types of the magnetic space groups are defined as follows:
\begin{enumerate}
    \item[Type 1.] $G_M=G$, $\Theta$ is not a symmetry operation of the lattice.
    \item[Type 2.] $G_M=G\times\{E, \Theta\}$,  $\Theta$ is a symmetry operation of the lattice.
    \item[Type 3.] $G_M=H+g'(G-H)$, where $H$ is an index-2 subgroup of $G$, and $g'$ is an antiunitary operation formed by $\Theta$ or $\Theta$ compound with translation.
\end{enumerate}
The third type is the so-called 2-color group, which is further divided into two subdivisions:
\begin{enumerate}
    \item[Type 3a.] $G_M=H+\Theta(G-H)$.
    \item[Type 3b.] $G_M=H+\Theta\bm{\tau}(G-H)$, where $\bm{\tau}$ is a translation contained in $G$ but not in $H$, which is a half lattice vector, such as $\bm{a}/2, (\bm{a}+\bm{b})/2,(\bm{a}+\bm{b}+\bm{c})/2$, and so forth.
\end{enumerate}
For Type 3b, there are two conventions for the magnetic unit cell, one using the lattice vectors of the corresponding non-magnetic lattice. We adopt the second convention, which treats the magnetic primitive cell as the unit cell.

Type 1 is the usual point and space group, with a total of 230 distinct groups. It is the most confusing partition since it can describe ferromagnets, ferrimagnets, and their limiting case of ferrimagnets, namely, compensated antiferromagnets, and even antiferromagnets, non-collinear magnets, with or without SOC. Such diversity causes many difficulties in applying Type 1 magnetic groups to real materials, especially for people who are not familiar with magnetic group theory. The root of this confusion lies in two conventional practices: First, we do not explicitly separate the cases with and without SOC; second, we misuse spin-inversion operations, that is, we not only allow $\Theta$ but also unitary operations $\mathcal{C}'_2$ to flip the spin. Note that symbol $'$ here denotes that the rotation axis is perpendicular to the spin orientation.

Type 2 magnetic groups are also known as gray groups, consisting of 230 distinct groups. These groups can only be applied to nonmagnetic materials or paramagnets. This is a natural result due to the existence of $\Theta$ symmetry. Type 2 magnetic groups are clearly defined and require no further restriction.

Type 3 is the usual 2-color or black-and-white group, and there are 1191 distinct groups of this type. Conventionally, in black-and-white groups, $\Theta$ serves as the symmetry operation that reverses the spin. Since $\Theta$ does not affect the spatial space, SOC is excluded from such a description. However, we usually do not restrict $\Theta$ to be the only symmetry operation that can flip the spin. Therefore, Type 3 magnetic groups can describe not only antiferromagnets and non-collinear magnets, but even ferromagnets. This is a problem, as ferromagnets have only one "color", but it can be resolved by a subdivision.

\begin{figure}[bt]
\includegraphics[width=0.48\textwidth]{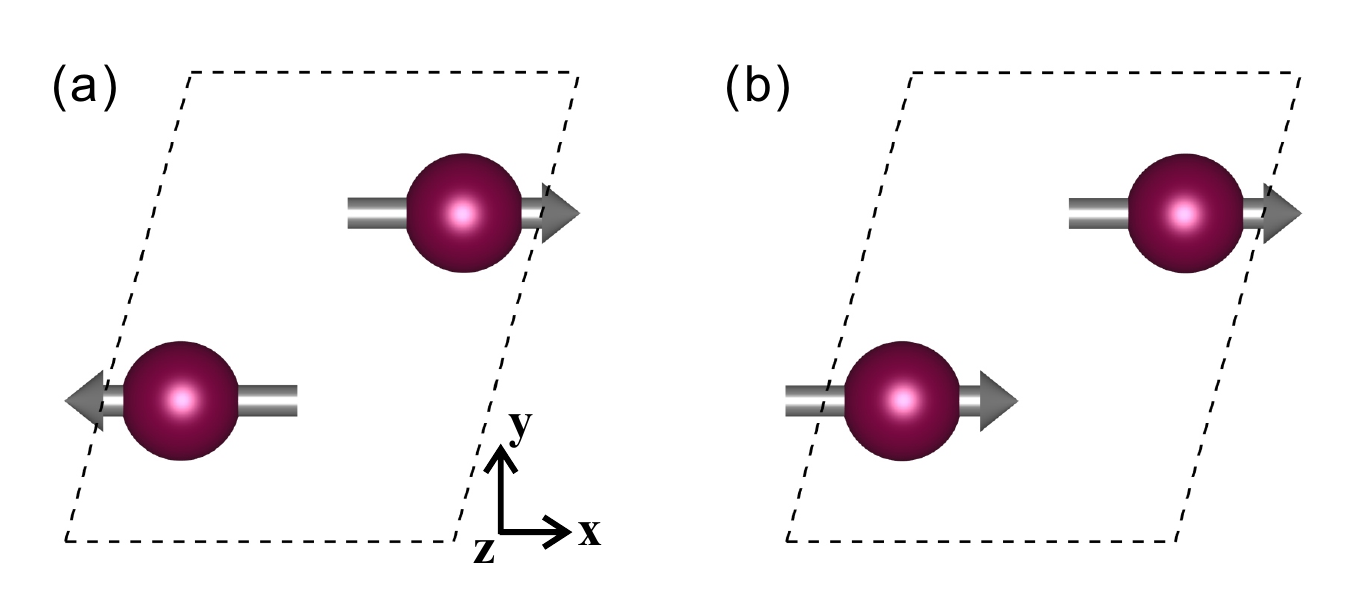}
\caption{\label{fig2} Illustrations of the confusion brought by conventional magnetic groups. (a) An antiferromagnetic lattice, with symmetry operations [$E, C_2$], which belongs to Type 1 magnetic groups. (b) A ferromagnetic lattice, with symmetry operations [$E, C_2\Theta$], which belongs to magnetic group Type 3. }
\end{figure}

The non-trivial subdivision of Type 3 magnetic groups\textemdash Type 3a is consisted of 674 so called lattice-equivalent magnetic groups. For a magnet whose symmetry group belongs to Type 3a, its magnetic unit cell is equivalent to its lattice unit cell. Here, the lattice unit cell refers to that of the corresponding non-magnetic lattice. Type 3a magnetic groups can describe antiferromagnets, non-collinear magnets, and even ferromagnets. Altermagnets without SOC also fall into this partition, and which will demonstrate later that altermagnets can coexist with SOC, not limited to a spin-pace-group description \cite{PhysRevX.12.031042,liu2022spin,jiang2024enumeration,xiao2024spin,chen2024enumeration}. 

Type 3b is a trivial spin-wave construction of the Type 1 group from ferromagnets to antiferromagnets with the spin wave vector $\bm{q}$ being a half-reciprocal lattice vector such as $\bm{K_a}/2, (\bm{K_a}+\bm{K_b})/2,(\bm{K_a}+\bm{K_b}+\bm{K_c})/2$ and so forth. There are a total of 517 so-called class-equivalent groups. The difference between Type 3a and 3b is that Type 3a contains no symmetry operation of the form $\Theta\bm{\tau}$. Type 3b includes 517 distinct groups and can only describe antiferromagnets with spin-degenerate energy bands in momentum space. Adding Types 1-3 together, there are a total of 1651 distinct magnetic space groups.

\section[\label{sec.3}]{Multicolor groups: General framework}
Now we introduce the multicolor group framework for classifying solids by their symmetries. Like the conventional 2-color group, in the multicolor-group framework, each local magnetic moment of the atoms is denoted by different colors, a color exchange operation must simultaneously exchange the positions of the atoms. A simple $m$-color group has a symmetry operation $\mathcal{P}$ that can permute the colors; that is, all remaining colors can be generated by acting $\mathcal{P}$ on the starting color. The collection of such symmetry operations has a group structure isomorphic to an Abelian group $\mathbb{Z}_m$. $\mathcal{P}$ also has a real space and a spin space part as $\{R_i||R_{\beta}\}$, with requirement that the order of $R_i$ and/or $R_{\beta}$ is $m$. Generally, $\mathcal{P}$ is an extended $C_m$ rotation, while for 2-color groups, in addition to $C_2$, $R_{\beta}$ can be taken as $\Theta$. Therefore, the group structure of a simple $m$-color group $G^m$ can be written as a coset decomposition \cite{group1999}:
\begin{equation}
\label{pgroup}
G^m=H+\mathcal{P}H+\mathcal{P}^2H+\cdots\mathcal{P}^{m-1}H=H\times\mathbb{Z}_m(\mathcal{P})\ ,
\end{equation}
where $H$ is an index $m$ subgroup of the space group $G$ corresponding to the non-magnetic lattice. 
A complex multicolor group is composed of multiple $m$-color groups, which means that there exists another symmetry operation $\mathcal{A}$ of order $n$ that can exchange $n$ sets of colors. The group structure of a complex $n\times m$-color group $G^{n\times m}$ can be written on top of a simple $m$-color group as:
\begin{equation}
\label{npgroup}
    G^{n\times m}=G^m+\mathcal{A}G^m+\cdots\mathcal{A}^{n-1}G^m=H\times\mathbb{Z}_n(\mathcal{A})\times\mathbb{Z}_m(\mathcal{P})\ .
\end{equation}

\begin{table}
\renewcommand{\arraystretch}{1.8}
\centering
\caption{Symmetry group of monolayer V$_2$Se$_2$O. Here $C'_2$ refers to $C_x$ or $C_y$, $C''_2$ refers to $C_{xy}$ or $C_{\bar{x}y}$.}
\label{table1}
    \begin{tabular}{c|c|c||c|c}
    \hline
        $D_{4h}$ &  \multicolumn{2}{c||}{w/o SOC}  & \multicolumn{2}{c}{with SOC} \\
        \hline
      \multirow{6}{*}{\includegraphics[scale=0.25]{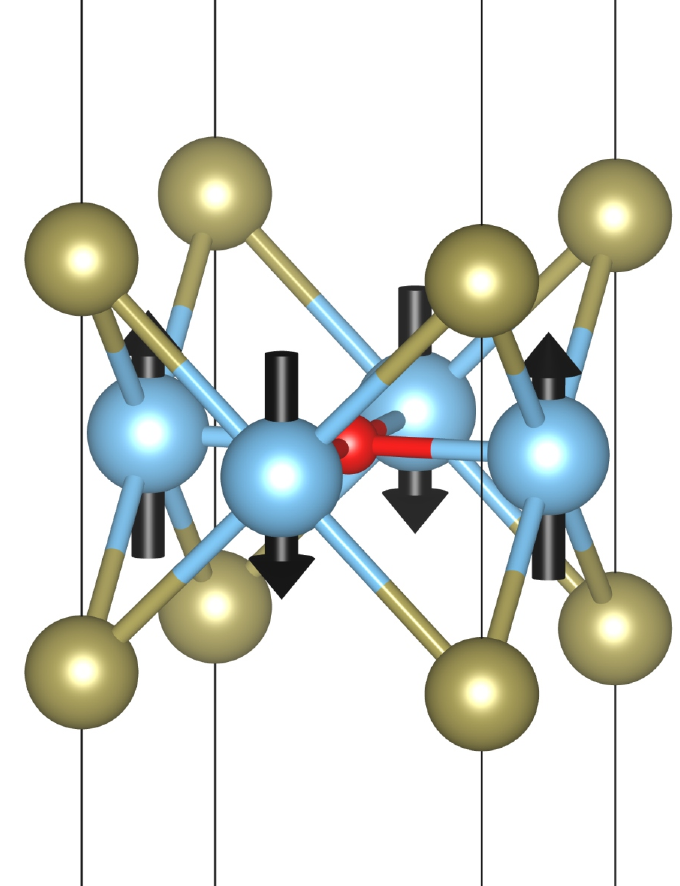}}  &  E            & 2$\Theta C''_2$    & E  & 2$\Theta C'_2$ \\ 
        &  $C_2$        &                        & $C_2$  &  \\ 
        &  i            & 2$\Theta C_4$      & i  & 2$\Theta C_4$ \\ 
        &  $\sigma_h$   &                        & $\sigma_h$  &  \\ 
        &  2$C'_2$      & 2$\Theta S_4$      & 2$C''_2$  & 2$\Theta S_4$ \\ 
        &  2$\sigma_v$  & 2$\Theta\sigma_d$ & 2$\sigma_d$  & 2$\Theta\sigma_v$ \\ 
       \hline
    \end{tabular}
\end{table}

\section[\label{sec.4}]{Multicolor groups: without SOC}
Having established the general framework of the multicolor group, the first set of questions one may ask is: there are 2-color groups in the conventional magnetic group, where are the 0-color, 1-color, and multicolor groups? What are multicolor groups for? We first restrict our analysis to collinear magnets. We further restrict $\Theta$ to be a unique operation that can reverse the spin. In comparison, crystal-symmetry operations such as $\mathcal{C}'_2$ act only in spatial space. Under these two restrictions, for collinear magnets without SOC, we have:
\begin{enumerate}
    \item[Type 0.] 0-color. $G_M=G\times\{E, \Theta\}$,  $\Theta$ is a symmetry operation.
\end{enumerate}
This group is nothing but the previously defined gray group, or colorless group. Such an identification is reasonable, as $\Theta$ does nothing to an object without any color.
\begin{enumerate}
    \item[Type 1.] 1-color. $G_M=G$, $\Theta$ is not a symmetry operation.
\end{enumerate}
Obviously, $\Theta$ cannot be a symmetry operation since there is only one color for the atoms. This type of symmetry group can be applied to ferromagnets, ferrimagnets, and the limiting case of ferrimagnets, namely, compensated antiferromagnets. The aforementioned trivial multicolor group belongs to this type. This trivial symmetry group should be singled out, since many non-collinear magnets have no symmetry operation other than the identity. The application to compensated antiferromagnets and non-collinear magnets may seem to challenge the concept of the 1-color group. However, we emphasize that here each magnetic moment on an atom must behave independently, just like in a ferromagnet. There must be no symmetry operation that relates atomic sites with different local magnetic moments. Belonging to the 1-color group, a more suitable name for compensated antiferromagnets would be zero-moment ferrimagnets.
\begin{enumerate}
    \item[Type 2.] 2-color. $G_M=H+g'(G-H)$, where $H$ is an index-2 subgroup of $G$, and $g'$ is an antiunitary operation formed by $\Theta$ or $\Theta$ compound with translation.
\end{enumerate}
This 2-color group is equivalent to the traditional Shubnikov space group, and can only be applied to antiferromagnets. 

\begin{table}
\renewcommand{\arraystretch}{1.8}
\centering
\caption{Symmetry group of NdB$_2$C$_2$. }
\label{table2}
    \begin{tabular}{c|c|c||c|c}
    \hline
        $D^5_{4h}$ ($\bm{\tau}=[1/2, 1/2, 0]$) &  \multicolumn{2}{c||}{w/o SOC}  & \multicolumn{2}{c}{with SOC} \\
        \hline
      \multirow{6}{*}{\includegraphics[scale=0.3]{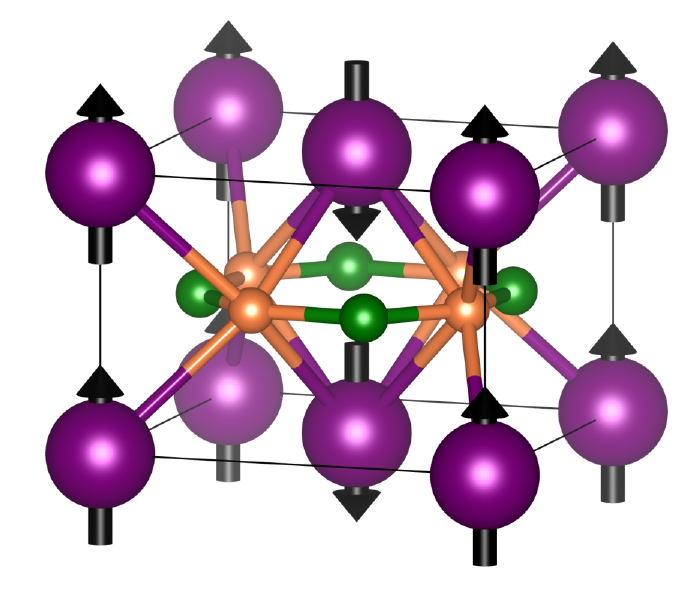}}  &  E            & 2$\Theta\bm{\tau}C'_2$    & E  & 2$\bm{\tau}C'_2$ \\ 
        &  $C_2$        &                        & $C_2$  &  \\ 
        &  i            & 2$\Theta\bm{\tau}C''_2$      & i  & 2$\bm{\tau}C''_2$ \\ 
        &  $\sigma_h$   &                        & $\sigma_h$  &  \\ 
        &  2$C_4$      & 2$\Theta\bm{\tau}\sigma_v$      & 2$C_4$  & 2$\bm{\tau}\sigma_v$ \\ 
        &  2$S_4$  & 2$\Theta\bm{\tau}\sigma_d$ & 2$S_4$  & 2$\bm{\tau}\sigma_d$ \\ 
       \hline
    \end{tabular}
\end{table}

As a demonstration, we consider a few examples to illustrate the advantage of this multicolor group classification. By using these examples, we aim to eliminate the confusion caused by the conventional magnetic group classification. As shown in Fig.~\ref{fig2}, the two crystals have the same lattice symmetry, while the left one is an antiferromagnet, and the right one is a ferromagnet. Conventionally, the symmetry operations of the antiferromagnet are [$E, C_2$], which belongs to Type 1. While the symmetry operations of the ferromagnet are [$E, C_2\Theta$], which belongs to Type 3a. It is confusing that a ferromagnet has a Type 3 symmetry group, a 2-color group, while an antiferromagnet has a Type 1 symmetry group, which is compatible with SOC. The root of this confusion lies in allowing unitary operations, other than $\Theta$, to also reverse the spin. Such confusion can be resolved by using the multicolor group classification. Under the restrictions induced by the 1-color and 2-color groups, the ferromagnet has the symmetry operations [$E, C_2$], which belong to the 1-color group, consistent with our classification scheme; while the antiferromagnet has the symmetry operations [$E, C_2\Theta$], which belong to the 2-color group, also consistent with our classification. Now the criterion for classification is clear, and there is no ambiguity. We no longer need to confuse symmetry groups with or without SOC.

\section[\label{sec.4}]{Multicolor groups: with SOC}
After demonstrating the efficiency of the multicolor group classification for solids without SOC, we now discuss solids having SOC. Without the SOC, a rotation of all the spins of the atoms results in an equivalent Hamiltonian. Therefore, a spin space group symmetry operation $\{R_i||R_{\beta}\}, R_i\neq R_{\beta}$ makes no difference from a symmetry operation $\{R_i||R_{\theta}\}$, where $\theta$ is an arbitrary rotation angle. This is a manifestation of the subspace symmetry \cite{ma2025subspace}.
With the SOC, a symmetry operation must affect both the spatial and spin degrees of freedom simultaneously and leave the magnetic structure invariant. In this case, both $\Theta$ and the unitary operations $\mathcal{C}'_2$ can flip the spin. The symmetry groups with SOC may be viewed as a special case of those without SOC, but this is not a simple deduction since the orientations of the local moments will influence the allowed symmetry operations. Generally, when the orientations of the local moments are along the principal rotation axis, the compound will exhibit the highest symmetry group, possibly having the same group order as that of the case without SOC; in other cases, the symmetry will be lower. 

For the 0-color group, the case with SOC can be well described by the established double group theory. For the 1-color group, introducing SOC imposes a strong restriction on the magnetic structure, the orientation of the spin can no longer be arbitrary. The detailed symmetry operations for a 1-color group, or a ferromagnet with SOC, can be diagnosed by a global axial vector, which corresponds to the net magnetic moment of the solid. Note that any symmetry group containing two independent $C_2$ rotations cannot leave the axial vector invariant, nor the magnetic structure. 
For the 2-color group, the definition still follows the form of the Type 3 magnetic group. However, the definitions of $H$ in Type 3a and 3b magnetic groups need to be modified, since now the unitary operations $\mathcal{C}'_2$ can also flip the spin. There are three subdivisions:
\begin{enumerate}
    \item[Type 2a.] $G_M=H+g'(G-H)$, where $H$ is an index-2 subgroup of $G$,and $g'$ is an antiunitary operation formed by $\Theta$ or $\Theta$ compound with translation. An example involving monolayer V$_2$Se$_2$O, which is an altermagnet that compatible with SOC\cite{ma2021multifunctional}, is shown in Table~\ref{table1}.
    \item[Type 2b.] $G_M=H+\mathcal{C}''_2(G-H)$, where $H$ is an index-2 subgroup of $G$, and $\mathcal{C}''_2$ is $\mathcal{C}'_2$ or its combination with translation. An example involving altermagnetic NdB$_2$C$_2$ \cite{Ohoyama2000} is shown in Table~\ref{table2}. Note that $\mathcal{C}'_2$ is not an antiunitary operation.
    \item[Type 2c.] This type is for non-collinear magnets. $G_M=H+\mathcal{P}(G-H)$, where $H$ is an index-2 subgroup of $G$, and $\mathcal{P}$ is either a $\mathcal{C}_2$ rotation coplanar with the two magnetic moments or a mirror reflection $\sigma$, possibly combined with a half-lattice translation $\bm{\tau}$.
\end{enumerate}
where Type 2c applies to non-collinear magnets. 

Starting from the 2-color group, there are spin-wave constructions. This spin-wave lattice is simply a result of magnetism, namely, without magnetism, the magnetic unit cell of a compound with a $m$-color group spin-wave structure is an enlargement of the nonmagnetic unit cell by $m$ fold along $\bm{q}$. The symmetry group is generated by $\bm{\tau}_mC_{m}$, and an example is the recently proposed $p$-wave magnets \cite{hellenes2023p,brekke2024minimal,sukhachov2024impurity,ezawa2024topological,maeda2025classification,chakraborty2025highly,soori2025crossed,nagae2025flat,sukhachov2025coexistence,yamada2025gapping,song2025electrical,zhou2025anisotropic}.

\begin{figure}[tb]
\includegraphics[width=0.48\textwidth]{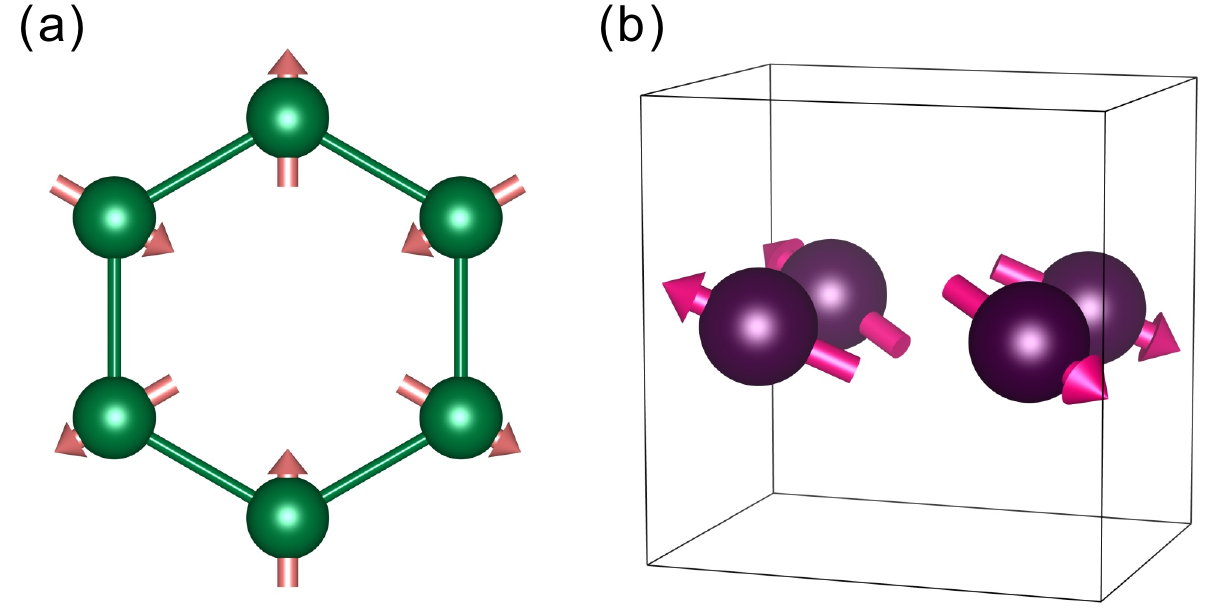}
\caption{\label{fig3} Illustrations of the $m$-color groups. (a) A simple 6-color group, generated by $C_{6z}\Theta$ or $S_{6z}$. (b) A complex 2$\times$2-color group, composed of i$\Theta\times C_{2x}$.}
\end{figure}

As stated above, for non-collinear magnets, SOC must be included. Therefore, in the following discussions, SOC is included. We first emphasize that, for crystals, $m$-color groups with $m>6$ are meaningless, because only 1, 2, 3, 4 and 6-fold rotations are allowed. Since we have discussed $m = 2$, here we only need consider $m = 3, 4$, and 6, which correspond to non-collinear magnets where SOC is present. For a simple $m$-color group with $m\geq3$, beside identity E and inversion i, all other symmetry operations must be generated by a $C_m$ rotation. We have:
\begin{enumerate}
    \item[Type 3.] The color exchange operations of a 3-color group are generated by a $C_3$ rotation or its composition with color-preserving operations such as fractional translation $\bm{\tau}$, or inversion i. The complete symmetry group has the form of Eq.~\ref{pgroup}, which can be decomposed into three disjoint cosets according to the $C_3$ rotation. The 3 local moments of magnet can be either coplanar or non-coplanar.
    \item[Type 4.] For the 4-color group with even index, the color exchange operations can be more interesting. Other than the color-preserving operations, color exchange symmetry operations $\mathcal{P}$ with index 2, such as $\Theta$ and mirror reflections, can also be composed with the principal $C_4$ rotation.
    \item[Type 6.] The 6-color group is analogous to the 4-color group. Its color exchange operations are generated by a $C_6$ rotation, or its composition with some color-preserving operations such as fractional translation $\bm{\tau}$, inversion i, or with color exchange symmetry operations $\mathcal{P}$ such as $\Theta$ or mirror reflections. An example is shown in Fig.~\ref{fig3}(a).
\end{enumerate}

Interestingly, other than the simple $m$-color groups, there also exist complex multicolor groups. A complex multicolor group, which contains two sets of color exchange symmetry operations, can be written as an $n\times m$-color group. It can be considered as having either $n$ sets of $m$-color groups or $m$ sets of $n$-color groups, other than 1 set for the general simple $m$-color groups. Since rotations about different axes do not commute, in crystals we only have $2\times m$-color groups with two sets of simple $m$-color groups, as follows:
\begin{enumerate}
    \item[Type 4a.] A $2\times 2$-color group has four colors but it cannot be generated by a single Abelian symmetry operation like the usual 4-color groups. It can be either coplanar or non-coplanar. A coplanar $2\times 2$-color group is simply composed of two collinear 2-color groups, while there is no general $C_4$ rotation that connects the two sets of symmetry groups. An example of a non-coplanar $2\times 2$-color group is shown in Fig.~\ref{fig3}(b).
    \item[Type 6a.] A $2\times 3$-color group is similar to the $2\times 2$-color group, consisting of two sets of 3-color groups. The complete symmetry groups are the direct product of the two sets of color exchange symmetry operations.
    \item[Type 8.] Like the $2\times 3$-color group, a $2\times 4$-color group is made of two sets of 4-color groups.
    \item[Type 12.] Similarly, a $2\times 6$-color is made of two sets of 6-color groups.
\end{enumerate}
Thus, we have established the framework of multicolor group classification for molecules and solids, as summarized in Table~\ref{table3}. However, some points remain open to discussions. First, the restriction $m\leq 6$ is induced by translational symmetry; for molecules, we can have $m>6$ multicolor point groups. Second, without the SOC, available spin-wave construction within the multicolor groups are also abundant, including having $m>6$. Note that in Ref.\cite{jiang2024enumeration}, more than 10000 distinct spin space groups have been enumerated.

\section[\label{sec.5}]{Outlooks}
As a final remark, human understand the world through classification, for example, organisms are classified into animals, plants, and microorganisms. We always seek a unified framework to classify objects, known or unknown. However, as our knowledge of the universe deepens, older frameworks may become insufficient, and new classification schemes or redefinitions become necessary. This motivates us to develop the multicolor group classification framework, extending beyond conventional magnetic group theory, to classify all molecules and solids by their symmetries, whether nonmagnetic or magnetic. Just like magnetic space groups, whose importance goes beyond merely describing magnetic order—they also serve as key tools for predicting and constraining electronic band structures, optical responses, magnetoelectric couplings, symmetry-protected topological phases, and spintronic devices. We believe that the multicolor group classification here will offer a unified language for exploring symmetry, topology, and complex quantum phenomena in molecules and solids, and will mark a further step toward the symmetry classification of matter that is not only limited to magnets, but also to ferroelectrics, multiferroics, and beyond.

\begin{acknowledgments}
We thank Dr. Y. Zhao for helpful discussions and comments. H.-Y. Ma thanks the financial support from  National Natural Science Foundation of China (Grant No. 12504225), Guangdong Provincial Quantum Science Strategic Initiative (GDZX2401001).
J.-F. Jia thanks the Ministry of Science and Technology of China (Grants No. 2019YFA0308600, 2020YFA0309000),  NSFC(Grants No. 92365302,  No. 22325203, No. 92265105, 92065201, No. 12074247, No. 12174252), the Strategic Priority Research Program of Chinese Academy of Sciences (Grant No. XDB28000000) and the Science and Technology Commission of Shanghai Municipality (Grants No. 2019SHZDZX01, No. 19JC1412701, No. 20QA1405100) for financial support. J.-F. Jia thanks the financial support from Innovation program for Quantum Science and Technology (Grant No. 2021ZD0302500).
\end{acknowledgments}

\bibliography{refs}

\end{document}